\documentclass[11pt,twocolumn,letterhead]{article}
\usepackage[T1]{fontenc}
 \pdfoutput=1
\usepackage{ntheorem}
\theoremseparator{:}
\newtheorem{hyp}{RQ}
\newtheorem{hypo}{H}

\usepackage{cvpr}
\usepackage{times}
\usepackage{epsfig}
\usepackage{graphicx}
\usepackage{amsmath}
\usepackage{amssymb}
\usepackage{array}
\usepackage{textgreek}
\usepackage[font=footnotesize,labelfont=bf]{caption}
\usepackage[margin=0.8in]{geometry}

\usepackage[pagebackref=true,breaklinks=true,letterpaper=true,colorlinks,bookmarks=false]{hyperref}

\cvprfinalcopy 

\def\cvprPaperID{****} 

\ifcvprfinal\fi

\begin{document}

\title{Forking Around: Correlation of forking practices with the success of a project}

\author{Anurag Dhasmana\\
{\tt\small adhasmana@ucdavis.edu}
\and
Arindaam Roy\\
{\tt\small aroy@ucdavis.edu}
\and 
Divjeet Singh Jas\\
{\tt\small djas@ucdavis.edu}
\and
Kiranpreet Kaur\\
{\tt\small kaykaur@ucdavis.edu}
\and 
Pinn Prugsanapan\\
{\tt\small prugsa@ucdavis.edu }
}
\maketitle

\begin{abstract}
\vspace{-0.5em}

Forking-based development has made it easier and straightforward for developers to contribute to open-source software (OSS). Developers can fork an existing project and add changes in their local version without interrupting the development process in the main project. Despite the efficiency of OSS, more than 80\% of the projects are not sustainable \cite{sustainability_forecast}. Identifying the elements related to OSS success can enlighten developers regarding the sustainability of a project. In our study, we explore whether or not the inefficiencies which arise due to forking-based development like redundant development, fragmented communities, lack of modularity, etc. \cite{what_the_fork} have any relation to the outcome of a project in terms of sustainability. We formulate eight metrics to quantify attributes for projects in the ASFI dataset \cite{dataset}. To find the correlation between the metrics and the success of a project, we built a logistic regression model to metrics with significant p-values and performed backward stepwise regression analysis, using the stepAIC function in R to cross-check our findings. The findings show that modularity, centralized management index, and hard forks are consequential for the success of a project. Developers can use the outcomes of our research to plan and structure their projects to increase the probability of their success.

\vspace{-0.6em}
\end{abstract}

\section{Introduction}
OSS is freely available to users, who can modify, enhance or view the source code implementation. The massive popularity of OSS is due to its collaborative nature. Collaboration is crucial for developing software projects at scale. Developers around the globe come together to build an application or add functionality as a team and sharpen their skills.  Collaboration helps in engendering novel ideas from the community of diverse developers, produces better outcomes, and envisions support for a project at a massive scale \cite{collaboration}.

The birth of social coding sites such as GitHub has completely democratized the concept of collaboration in OSS. Before the existence of git, developers had limited options for collaborative development such as SVN. Developers can nowadays fork a project and add their changes to their local copy without needing approval from the project owners.

The idea of forking in OSS has evolved. Traditionally, forking implied cloning a project and branching off new independent development. Forking was rare and was often intended to compete with or supersede the original project \cite{how_to_run, hackers_on_forking, perspective_on_forking}. In the modern view, forks in version control systems such as git are public copies of repositories that developers can modify, potentially, but not necessarily, to integrate those changes back into the original repo. Forking has made the reuse of existing code of software projects more straightforward. Instead of exchanging emails with software patches, developers can separately modify and experiment with the original project to either start a new project or contribute to an existing one \cite{why_fail}. 

With the rise of social coding and explicit support in version control systems, forking of OSS repositories has become very popular. Consequently, developers are creating open-source code at an unprecedented rate. GitHub has more than 19 million users and 52 million repositories (without excluding forks) \cite{why_fail}. 

Despite the popularity of OSS and the relative ease of development provided by forking-based social coding, more than 80\% of the OSS projects fail \cite{sustainability_forecast}. \textbf{This great failure rate makes us ponder from a social point of view: Why do most of the OSS projects do not succeed?} According to our definition, success means self-sustaining in the longer run.  Previous research has attempted to solve the question mentioned above \cite{why_fail, collaboration}. However, to the best of our knowledge, no work has been done to investigate a direct relationship between an OSS project's success or failure and forking practices.

We make a foray in this direction as developing projects with forks comes with costs. Independent development means contributions are not always visible to others unless there is an explicit merge-back attempt. Coordination overhead rises as the number of forks increases \cite{feature_based}. Specifically, we notice the following hurdles:
\begin{enumerate}
    \item Redundant Development: Contributors often independently implement similar functionality in their local forks that wastes development time and resources \cite{feature_based, what_the_fork}.
    \item Fragmented Communities: Developers maintain many software versions in parallel through their forks, resulting in fragmented communities. It makes it harder for users to choose the project variant that best matches their needs or is the most actively maintained \cite{what_the_fork}. Hard forks, sometimes known as "forks of forks," push the project in different directions than intended by the core team.
    \item Management Issues: Lack of modularity in the code structure and limited communication between contributors and core team members impede the growth of a project. Monolithic architecture is difficult to reason with and even worse to change. Inadequate communication between contributors and the core team leads to uncertainty about future growth.
    \item Scalability Concerns: As the project's size grows, so does the number of commits and the frequency of pull requests. When a significant number of commits and pull requests in a short period, it becomes difficult for the maintainers to assess all of the changes promptly.
\end{enumerate}

Overall, we consider many forking-based development inefficiencies in projects; specifically redundant development, fragmented communities, lack of modularity, insufficient communication between developers and contributors, pushing large. numbers of commits, and frequently pull requests (PRs) to the main branch. We investigate how these inefficiencies are co-related to a project's success or failure. Comprehensively, we aim to establish a weak causal relationship between efficient forking practices and the success of a project. Specifically, \textbf{Does adhering to the efficient forking practices by a project correlate with its success?} We explore further to find answers to the following questions:
\begin{hyp}\label{rq:first}
Is there a difference in the number of duplicate pull requests (PRs) and hard forks causing redundancy between successful and unsuccessful projects?
\end{hyp}
\begin{hyp}\label{rq:second}
Does the frequency of pull requests (PRs) raised to the main project have anything to do with the success of a project?
\end{hyp}
\begin{hyp}\label{rq:third}
Is the size of the commits pushed to the main branch relevant to the success of a project?
\end{hyp}
\begin{hyp}\label{rq:fourth}
Is the modularity of a project characteristic of its success?
\end{hyp}
\begin{hyp}\label{rq:fifth}
 Is coordination between the developers and the core team when working on a problem or a new feature important for a project's success?
\end{hyp}

To collect our metrics for quantifying the concerns which engender due to the forking-based inefficiencies and find their significance to the success and failure of a project, we used the ASFI dataset \cite{dataset}. The dataset contains a categorized list of retired (failed) and graduated (successful) projects with their corresponding GitHub URL. We collect the set of metrics representing forking-based attributes using GitHub APIs and python library, PyDriller. Next, we build a logistic regression model to evaluate their significance to a project's success. We found that the central management index, logic coupling index, additive contribution index, and presence of hard forks are significant metrics that correlate with a project's success.

We summarize the contribution of our research as follows:
\begin{enumerate}
    \item Investigation of various inefficiencies emerging due to forking-based development in OSS and defined those quantitatively through a set of metrics.
    \item Collection of data for eight metrics for further analysis using over 300 projects mentioned in the ASFI dataset. 
    \item Analysis of metrics using a logistic regression model and AIC to find the significant metrics representing the forking-based development practices affecting the success of a project.
\end{enumerate}

The code for the metric calculation and analysis is available on GitHub \footnote{Data and scripts are available: \url{ https://GitHub.com/anurag-asu/se-project-team-17}}.

\section{Literature Review}
In this section, we shed light on the current research and techniques for analyzing projects that use code forking. We also discuss some recent project sustainability work. We will present various methods and their shortcomings as they correlate to our work.

{\bf Apache Software Foundation Incubator:} Open and diverse communities tend to be more robust and productive than closed ones \cite{apache}. Apache Software Foundation Incubator (ASFI) is one of the open communities that gives guidelines on how to accept new committers and enables more community development. They believe that good software emerges if they foster and take good care of the community \cite{sustainability}. 

One of the goals of ASFI is to help projects become self-sustainable and eventually join the ASF community. ASF guidelines require all projects to make all communication publicly available through their mailing lists and committers are required to discuss first before making any changes. If it did not happen on the mailing list, it did not happen \cite{oss_mailinglist}. One unique aspect of ASF is that these projects are incubated during their “nascent stage of development,” when there is continuous recruitment of new committers, and detailed discussions are happening with project mentors \cite{dataset}. This data is collected until the project exits the incubator. Since the definition of success is entirely subjective, categorizing projects as successful or not is quite challenging \cite{collaboration}. However, ASF alleviates this challenge by employing experienced personnel to characterize the projects in the incubator. 

Yin et al. \cite{dataset} parsed the Apache Software Foundation Incubator (ASFI) to form a comprehensive dataset of 312 projects, classified as graduated, retired, and in incubation by the experienced admins of the AFSI community. This extrinsically labeled dataset forms our basis for categorizing successful and failed projects, as rightly envisioned by the authors.

{\bf Success and Sustainability of OSS Projects:}
Recently, ample research has been done on predicting the success of open source projects \cite{factors_oss, attributes_oss, determinants_oss} and finding the reasons for their failures \cite{why_fail}. From the research done, it seems that many projects fail due to their "one-man" nature \cite{collaboration}. Since project knowledge is limited to one or a few developers, projects will start to dwindle or die as soon as those single developers quit working. The phrase "Truck Factor" refers to the number of developers who must be hit by a truck before the project begins to suffer \cite{truck_factor}. Coelho et al. \cite{why_fail} presented a list of nine reasons like the project was usurped by a competitor, the project became functionally obsolete, lack of time of the main contributor, etc. as to why open source projects fail. The authors also demonstrated the presence of a significant gap in terms of optimal open-source maintenance, like the availability of contribution guidelines and the use of continuous integration, between unsuccessful projects and the most popular and active projects on GitHub. Other works attempted to develop a model based on statistical correlation \cite{planning_to_mature, measuring_oss_success, factors_oss}, but they did not have any objective heuristics to define the success of a project.

Identifying the elements related to OSS success can aid in the development of solutions when a project has a setback. In the paper \cite{sustainability}, the authors try to forecast a project becoming sustainable using their novel interpretable model. Also, based on these models, they offer an approach for real-time monitoring and actionable suggestions that projects may utilize to rectify their sustainability trajectories. While they help ongoing projects on the path of sustainability, our research will help developers to employ efficient forking and development practices from the inception of the project to increase its probability of being sustainable.

{\bf Forking-based Development:} With the increase in social coding, there is a rapid increase in forking, which is splitting an independent development line to foster more collaborative development. Prior research has also shown that forking increases community interaction opportunities \cite{social_coding_forking, community_engagement, pull_based_development, open_collaboration_public_sector}. For example, forks account for more than half of the commits in the Marlin project \cite{marlin}. The number of forks a project received seems to be a good reflection of its quality and worth as it indicates that a project had a lot of interest from the community and was therefore likely to be good or interesting \cite{social_coding_forking}. The increase in forks, however, spawned new challenges.

\emph{Redundant Development:} Many projects reject PRs because of redundant development \cite{what_the_fork, community_engagement, pull_based_development, almost_there, duplicatePR_dataset}. Working on a change only to learn that other developers have already made similar changes or have done so in tandem can be demotivating. To tackle this, many researchers have built software to detect such duplicates in the past \cite{feature_based, duplicatePR_dataset, detecting_duplicatePRs, identifying_redundancies}. In our research, we also develop a method to detect duplicates and calculate the ratio of duplicate PRs across different projects to investigate the effects of this inefficiency in a project. 

\emph{Fragmented Communities:} Many OSS projects attract independent contributors through forking. These forks can further draw more developers, creating a separate line of development. However, these hard forks can steer the project away from its original direction \cite{social_forking, marlin}. Such fragmentation might pose a threat to the long-term viability of open-source projects when two distinct communities form \cite{what_the_fork}. When such types of communities spawn, fragmentation-related attributes can hinder the development of the main project. We calculate the presence of hard forks across projects to explore if this attribute leads to a project’s collapse.

\emph{Modularity:} In an OSS development environment, modularity is a desirable attribute of projects \cite{factors_oss, oss_innovation, linux_edge}. MacCormack et al. \cite{exploring_complex_software} proposed that more modular projects would be more appealing to potential contributors. Midha et al. \cite{factors_oss} dis-covered that the modularity of open-source projects is positively related to technical success. However, modularity may not necessarily correspond to how developers think \cite{perspective_on_forking}. Researchers have presented various metrics for measuring project modularity in the past. For example, many methodologies use program analysis to identify relationships between software structures \cite{measure_functional_cohesion, discriminant_merit}. In our research, we employ some of these metrics, specifically, logic coupling index and additive contribution index, to assess the modularity of projects.

\emph{Coordination mechanisms:} OSS initiatives involve a large number of individuals (over hundreds), making coordination a very crucial aspect of a project. Bird \cite{bird2011sociotechnical} tried to find out if OSS projects are haphazard or organized. The author concluded that during the OSS life cycle, subgroups form. The same group of developers ends up working on the same files. The author, however, did not relate the coordination of developers to its sustainability. Zhou et al. \cite{what_the_fork} formulated two different metrics to quantify coordination among developers in a fork-based development. In our study, we employ both these metrics, namely centralized management, and pre-communication index, to find the measure of coordination.

\section{Hypotheses}

Our goal is to find the correlation of efficient forking with a project's success. Here we generate our hypotheses based on prior work and formulate metrics that will help us answer our research questions \ref{rq:first}, \ref{rq:second}, \ref{rq:third}, \ref{rq:fourth}, \ref{rq:fifth}. 

Many secondary forks (also known as forks of forks) contribute to other forks but not to the original repository, causing these forks, also known as hard forks to wander away from the main development line \cite{social_forking, concepts_operations}. Hard forks are seen as risky and antisocial to projects, as they can fragment a community and lead to confusion for both maintainers and users \cite{how_to_run, understand_open_source, hackers_on_forking, cathedral}. There aren't many examples of communities from the main project surviving a hard fork \cite{what_the_fork}. Redundant development, on the other hand, is one of the common reasons for rejecting PRs that impede contribution \cite{community_engagement, almost_there, duplicatePR_dataset}. Working carefully to resolve an issue only to discover that other developers have made similar improvements previously or in parallel can be demotivating. Zhou et al. \cite{what_the_fork} consider projects that reduce accidental redundancies and have fewer hard forks as more efficient. Thus, we hypothesize that a successful project should have a relatively fewer number of duplicate PRs and even fewer hard forks.
\begin{hypo} \label{hypo:first}
Ratio of duplicate pull requests are lower for a successful project
\end{hypo}
\begin{hypo} \label{hypo:second}
Presence of hard forks are lower for a successful project
\end{hypo}

Many researchers and practitioners emphasize the importance of modularity in open-source development \cite{oss_innovation, factors_oss, linux_edge}. Torvalds \cite{linux_edge} states, “for without [modularity], you cannot have people working in parallel.” MacCormack et al. \cite{exploring_the_structure} suggest that more modular projects can be more attractive to potential contributors. We, therefore, hypothesize that modularity is one of the factors that contribute to the success of a project. Logic coupling index and additive contribution index \cite{what_the_fork} are both predictors of modularity. A lower logic coupling index, which measures the degree to which two or more files change together, signals higher modularity. The additive contribution index, which measures to what degree contributions are additive, indicates better modularity for a higher value. 
\begin{hypo} \label{hypo:third}
Logic coupling index is lower for a successful project
\end{hypo}
\begin{hypo} \label{hypo:fourth}
Additive contributing index is higher for a successful project
\end{hypo}

Researchers have exhaustively explored different degrees of coordination and their tradeoffs in distributed collaboration. Brandts and Cooper \cite{centralization} state that central coordination makes it easier to plan and manage a product. We hypothesize that project coordinating contributions upfront encourages more focused development activities in a project that aligns with the maintainer’s vision. The issue tracker in version control systems keeps track of forks that contain the corresponding code changes for each issue, as other forks are of little interest to maintainers and unsolicited PRs remain ignored for years \cite{what_the_fork}. We also hypothesize that better pre-communication, i.e. evaluating developers' discussions on changes for resolving an issue with the core team before submitting PRs, leads to better coordination and, as a result, a higher project success rate.
\begin{hypo} \label{hypo:fifth}
Central Management index is higher for a successful project
\end{hypo}
\begin{hypo} \label{hypo:sixth}
Pre-communication index is higher for a successful project
\end{hypo}

A commit represents the snapshot of a project in a given state and has all the information about the code changes compared to the previous version. Prior work \cite{commit_size} has modeled the distribution of commit sizes in open source projects to find its relation with project sizes. The findings reveal that when the number of developers on a project grows, the size of commits reduces insignificantly. To correlate the effects of commit size with the success of a project, we use the dimensionality metric. It calculates the average number of lines of code changed per commit in a project. We, therefore, hypothesize that the size of the commits should not be a significant factor for a project's success.
\begin{hypo} \label{hypo:seventh}
Dimensionality index does not have a correlation  success of a project
\end{hypo}

OSS projects are developed globally by contributors, who collaborate through the pull-based model that lowers the barrier to entry for OSS developers by synthesizing, automating, and optimizing the contribution process \cite{duplicate_prs_empirical}. Hu et al. \cite{multi_review} have shown that more review-switching may bring longer resolution latency of a pull-request. As the number of PRs within a time frame increases, denoted by the frequency index, there will be longer resolution time and correspondingly fewer contributions to the project. Therefore, we hypothesize that the rate of contributions should have a relation with the project's capability to incorporate changes efficiently and be sustainable.
\begin{hypo} \label{hypo:eigth}
Frequency index value correlates with the success of a project
\end{hypo}

\section{Methodology}

In this section, we explain the process to gather and calculate the metrics to quantify the features of the projects.

\subsection{Dataset}
We calculate the metrics for the projects in the ASFI dataset \cite{dataset}. The projects are labeled as graduated or retired by the ASFI community. As a virtue of this, we have an objective classification of the projects. The dataset contains 312 data points consisting of 204 graduated, 60 retired, and 48 `in incubation` projects.
 
\emph{Data Pre-processing:} Out of these 312 projects, the GitHub URLs for 91 projects were incorrect or no longer functional as they returned HTTP 404 responses. We could only find 63 working URLs replacements for those 91 projects. A summary of the dataset used is given in the table [\ref{table:Dataset}] above.

\begin{centering}
\begin{table}
\small
\caption{\textbf{ASFI Dataset Summary}}
\small
\begin{tabular}{l*{3}{c}r}
             & Originally  & Working GitHub URLs\\
\hline
Incubating  & 48 & 48\\
Graduated  & 204 & 200 \\
Retired  & 60 & 36\\
\hline
Total  & 312 &  284\\
\hline
\end{tabular}\\
\label{table:Dataset}
\end{table}
\end{centering}

\subsection{Data Methods}
To gather the metrics for the projects, we used \textit{PyDriller} \cite{pydriller} and its' built-in functions to retrieve the commit data and data which is intrinsically present in the codebase. PyDriller, however, does not contain information about forks or pull requests. We used \textit{GitHub API} \cite{GitHubapi} to extract relevant information to calculate the metrics relying on the PRs, Issues, and comment data such as the presence of hard forks, ratio of duplicate PRs, etc. We used the \textit{pandas} library of python to interact with the data and collect the metrics. 

\textbf{Issues faced with GitHub:} GitHub API is free to use, however, its rate is limited to 5,000 authenticated requests per hour. As a result, certain metrics consumed days to run. To alleviate this issue, we had to batch and distribute the data gathering to multiple accounts. Some of the projects listed on the ASFI website are not hosted on GitHub. Though a GitHub mirror is present for some of these projects, these mirrors do not contain any kind of communication level data that one might expect from a GitHub project. This makes it ineffective to use such repositories as it contributes towards a threat to internal validity.

\textbf{Issues faced with PyDriller:} PyDriller is a slow tool and, through our experiments, we found it taking up to 24 hours to calculate metrics. This made it harder to iterate through code changes. 

\subsection{Metrics}

We calculate a set of eight metrics to find a correlation between efficient forking practices and the success of a project.
The metric collected has been summarized and visualized in Fig.\ref{fig:histogram}.

\textbf{Logic coupling Index:} is used to assess the modularity of the project by measuring the degree to which two or more files need to change together or co-evolve, in the most recent commits by contributors. We estimate the ratio of file pairs that were changed, i.e. added or modified, together in those commits out of all file pairs in the projects using PyDriller. Next, by computing the mean of recent commits, we aggregate this metric at the project level. We assess the most recent 150 commits for each project to remove bias from past but now altered practices. Lower logic coupling indexes suggest higher modularity, as fewer files are altered at the same time.
\textrm{Logic coupling Index =}
\[\textstyle\dfrac{\sum_{commit=1}^{commit=n} \textstyle\dfrac{\textrm{file pairs changed in the commit}}{\textrm{total file pairs in the project}}}{\textrm{n}} \]
\textrm{, where n = total number of commits}

\textbf{Additive contribution Index:} is a second modularity measure that assesses the extent to which contributions are additive, i.e contributions by adding files rather than editing existing ones. We calculate the ratio of new files added out of all files modified per commit, extracted through PyDriller. For a project, it’s the average ratio of overall commits from contributors. The higher additive contribution index shows that more modifications were additive, implying stronger modularity from the viewpoint of contributors.

\textrm{Additive contribution Index =}
\[\textstyle\dfrac{\sum_{commit=1}^{commit=n} \textstyle\dfrac{\textrm{number of files added in the commit}}{\textrm{number of files changed in the commit}}}{\textrm{n}} \]
\textrm{, where n = total number of commits}

\textbf{Ratio of Duplicate PRs:} is a clear metric for redundant work. We measure the fraction of the closed PRs rejected, by maintainers, stating as being redundant. We use GitHub API to fetch the attributes of the PR requests. To identify duplicate PRs in a project, we use regex matching and look for comments in the PR such as ‘duplicated’, ‘superseded’, ‘replicated’, etc. in PR comments.

\textrm{Ratio of duplicate PRs =}
\[\textstyle\dfrac{\textrm{duplicated closed PRs identified}}{\textrm{total PRs}} \]

\begin{figure}
    \centering
    \includegraphics[scale=0.25]{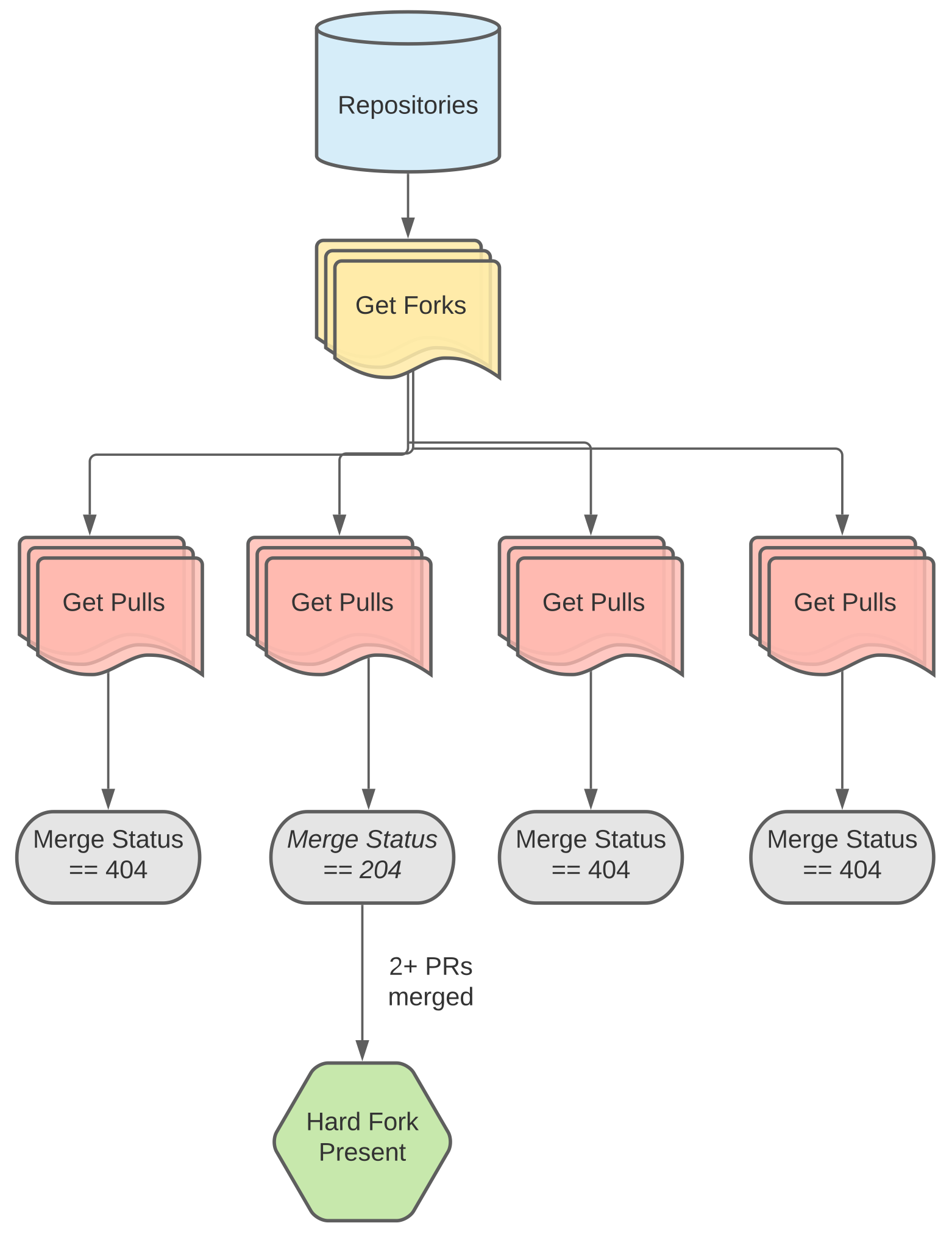}
    \small
    \caption{\textbf{Flowchart of finding hard forks using GitHub API calls. We first look at the number of forks for each repository. Then, we get the PRs for each fork, and check if the PR has been merged or not using the GitHub API "/merge" call for the PR. And, if 2 or more PRs have been merged successfully, we consider that a hard fork.}}
    \label{fig:hardfork}
\end{figure} 

\textbf{Presence of hard forks:} a fork that has at least 2 merged pull requests from external developers is identified as a hard fork \cite{what_the_fork}. For calculating this metric, as shown in Fig.\ref{fig:hardfork}, we use the REST API by GitHub to get a list of forks, and for each fork, we get a list of pull requests. Then, we see the merge status of each pull request in that fork to categorize if there are any merged PRs in a particular fork. If we find 2 or more merged PRs for a fork, we assign a boolean value of TRUE/FALSE for that project categorizing if the project has a hard fork or not.

\textbf{Centralized management Index:} This metric is used to quantify the contributions which start with a centralized discussion before a PR. It measures the fraction of PRs that link to issues out of all PRs from all contributors. The list of PRs is fetched using GitHub API for a project and regex is used to search for keywords like ‘issue \#123’ or ‘\#123’ in the title, comments, and/or git messages. Some projects refer to issues not hosted on GitHub. The reference, however, still follows the convention of starting with '\#'.

\textrm{Centralized Management Index =}
\[\textstyle\dfrac{\textrm{PRs with issues referenced in the title/description}}{\textrm{description total PRs}} \]

\begin{figure}
    \centering
    \hspace*{-1.6cm}
    \includegraphics[scale=0.6]{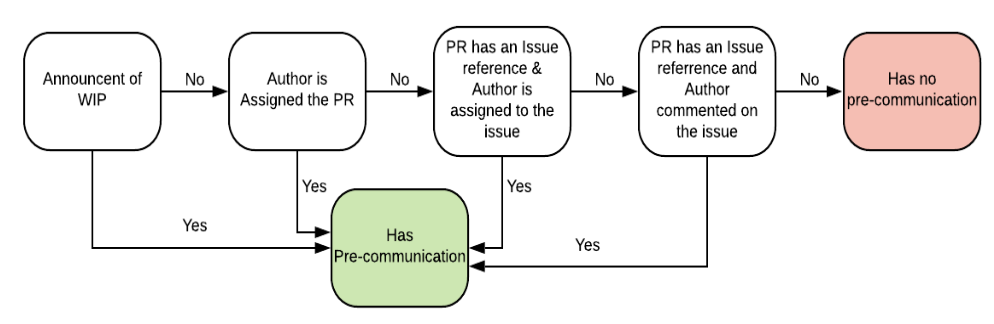}
    \small
    \caption{\textbf{Flow chart of figuring out if a PR’s author has pre-communicated about the issue/feature before starting to work on it. We first look for the announcement of ‘WIP’, then we look for if the author is one of the assignees of the PR. Next, we check if the PR is linked to an issue and parse the issue to check if the author has been assigned the issue. Finally, we check the comments in the linked issue for traces of participation by the author of the PR.}}
    \label{fig:pre_communication}
\end{figure}

\textbf{Pre-communication index:} This observes the coordination amongst the developers and contributors. in this metric, we look for the public announcement made by the contributors regarding the issue they are working on.
We do this in a hierarchical order, as shown in Fig.\ref{fig:pre_communication}. We first look for the announcement on the PR, then we look for if the author is one of the assignees of the PR. Next, we check if the PR is linked to an issue and parse these issues to check if the author has been assigned the issue or if the author has raised the issue. Finally, we check the comments in the linked issue for traces of participation by the author of the PR.

\textrm{Pre-communication Index=}
\[\textstyle\dfrac{\textrm{PRs with pre-communication}}{\textrm{Total PRs}} \]

\textbf{Dimensionality Index:} measures the size of commits, more precisely average lines of code changes (added + deleted) per file per commit in a project. This metric tries to capture the distribution of the size of commits between successful and failed projects.  We used PyDriller \cite{pydriller} and traverse all commits pushed by developers to a project to get the lines of code changed per commit. The dimensionality index value for a project is the average of lines of code changed to the number of files changed across all commits.

\textrm{Dimensionality Index =}
\[\textstyle\dfrac{\sum_{commit=1}^{commit=n} \textstyle\dfrac{\textrm{lines of code changed in the commit}}{\textrm{number of files changed in the commit}}}{\textrm{n}} \]
\textrm{, where n = total number of commits}

\textbf{Frequency Index:} is a metric to take into account the rate of contributions made by developers to a project and the degree to how difficult it is for the core team to incorporate the changes. It's measured as the average number of PRs raised to the main branch in a particular time interval. We decided on an initial time interval of 14 days for collecting this metric. Noticing a great variability in the data collected we then decided to check the number of PRs raised during a month. Since PyDriller does not have pull requests data we rely on GitHub APIs \cite{GitHubapi} to collect them for a project. Using paginated pull APIs we get all the PRs and count the number of PRs raised during 30 days. The index value is then just the average of the counts of all the PRs raised.

\textrm{Frequency Index =}
\[\textstyle\dfrac{\sum_{interval=1}^{interval=n}\textrm{total number of PRs raised}}{\textrm{n}} \]
\textrm{, where n = total number of intervals}

\begin{figure}
    \includegraphics[scale=0.21]{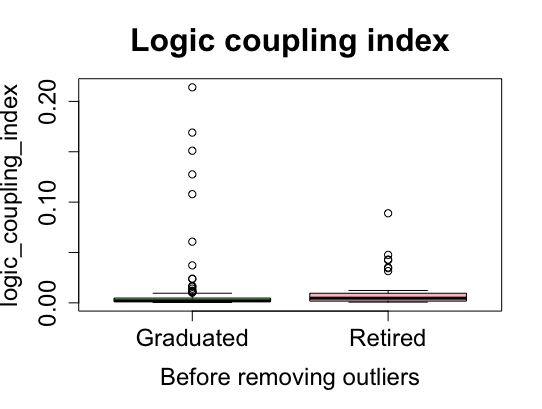}
    \small
   \label{fig:boxplot1}
    \includegraphics[scale=0.21]{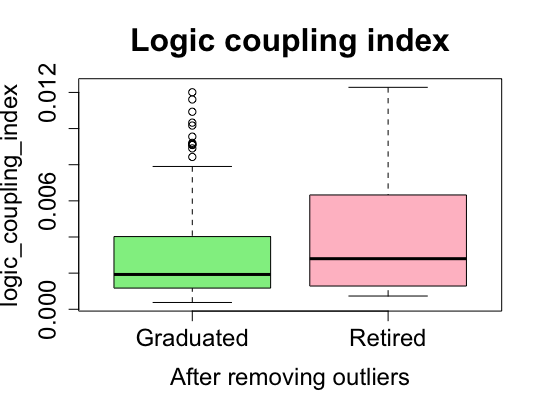}
    \small
   \label{fig:boxplot2}
    \includegraphics[scale=0.21]{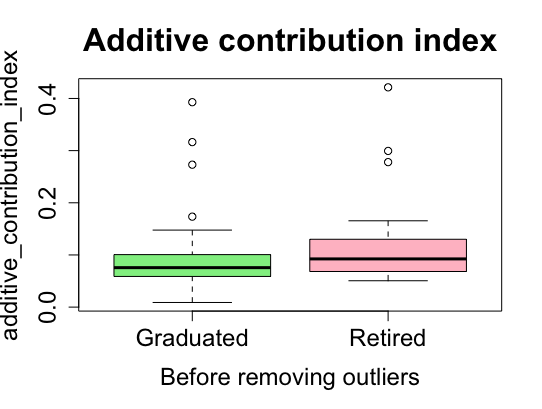}
    \small
   \label{fig:boxplot3}
    \includegraphics[scale=0.21]{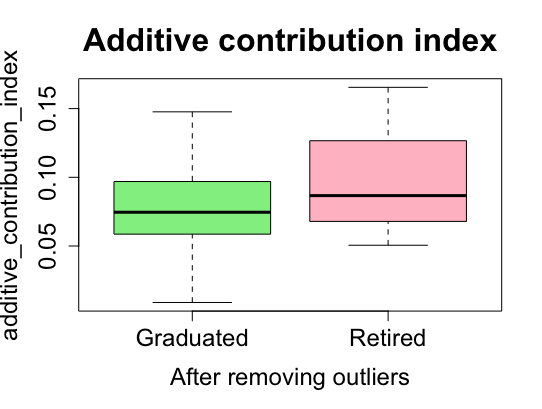}
    \small
   \label{fig:boxplot4}
    \includegraphics[scale=0.21]{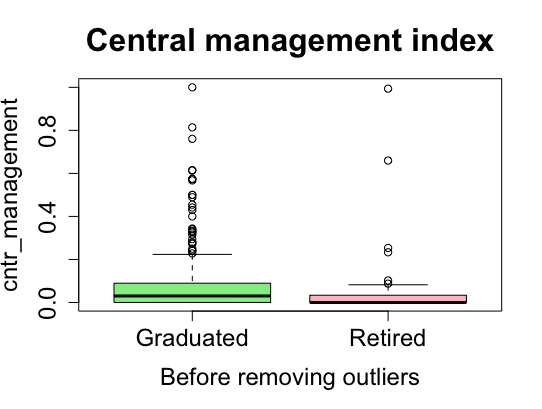}
    \small
   \label{fig:boxplot5}
    \includegraphics[scale=0.21]{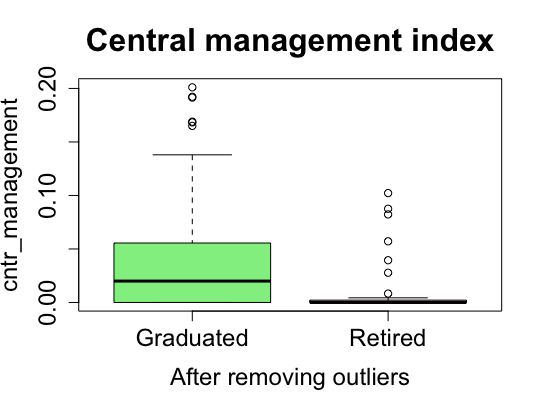}
    \small
   \label{fig:boxplot6}
   \caption{\textbf{Boxplot data distribution of significant metrics before (on the left) and after (on the right) removing outliers.\\ 
    Label: Green - Graduated projects, Pink - Retired projects.
    Note: Since \textit{presence of hard forks} has binary data, boxplot for it cannot be formed.}}
    \label{fig:boxplot}
\end{figure}

\subsection{Quantitative Analysis}
\begin{figure} 
    \includegraphics[scale=0.18]{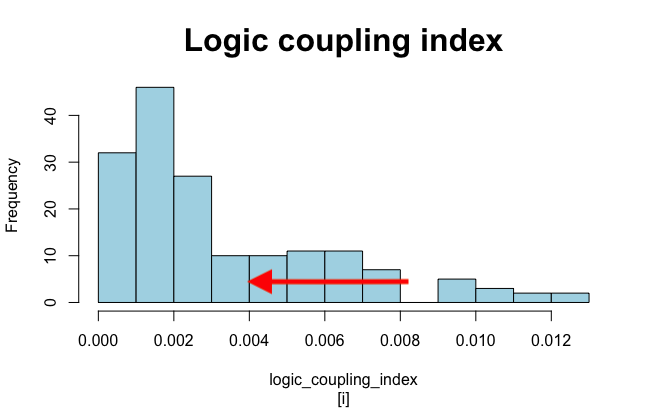}
    \small
    \label{fig:histogram1}
    \includegraphics[scale=0.18]{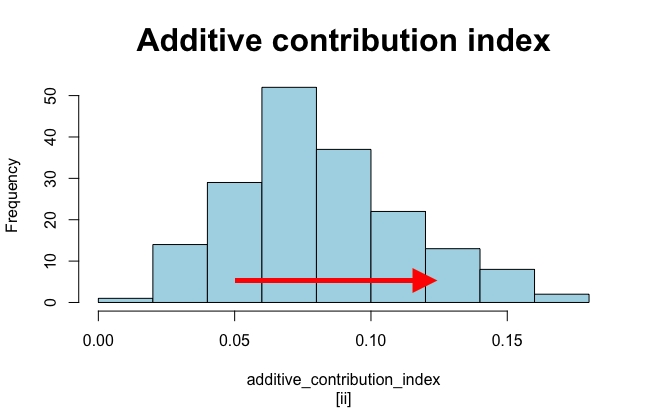}
    \small
    \label{fig:histogram2}

    \includegraphics[scale=0.18]{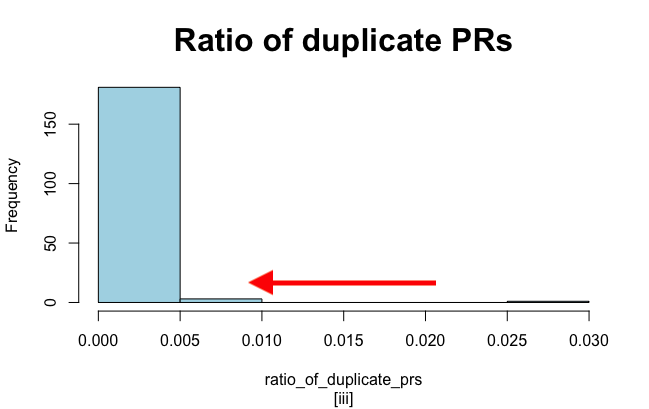}
    \small
    \label{fig:histogram3}
    \includegraphics[scale=0.18]{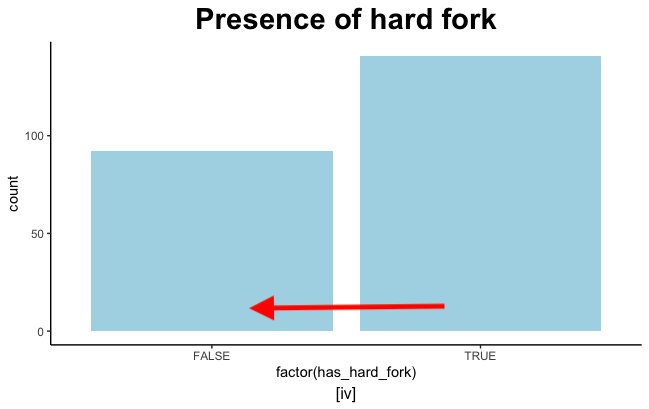}
    \small
    \label{fig:histogram4}
    
    \includegraphics[scale=0.18]{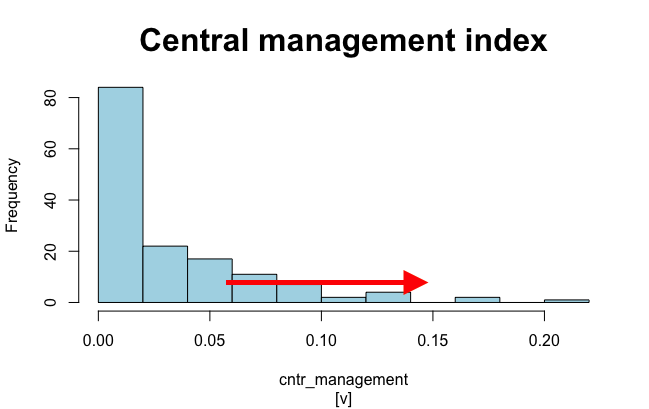}
    \small
    \label{fig:histogram5}
    \includegraphics[scale=0.18]{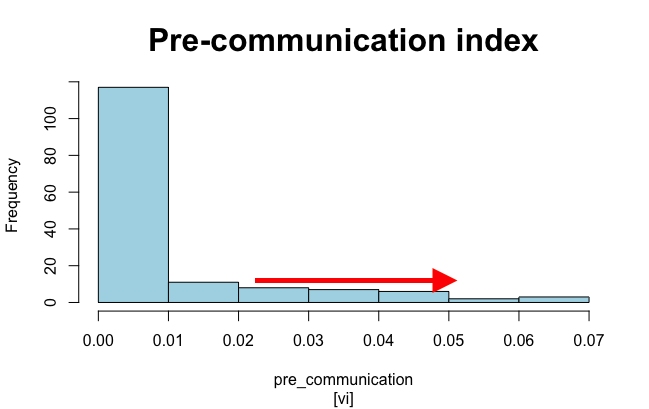}
    \small
    \label{fig:histogram6}
    
    \includegraphics[scale=0.18]{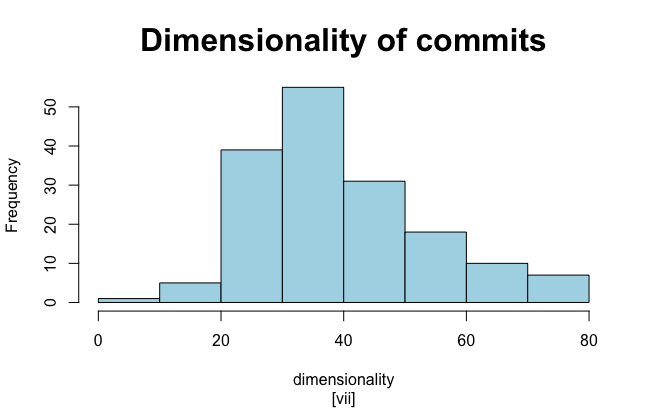}
    \small
    \label{fig:histogram7}
    \includegraphics[scale=0.18]{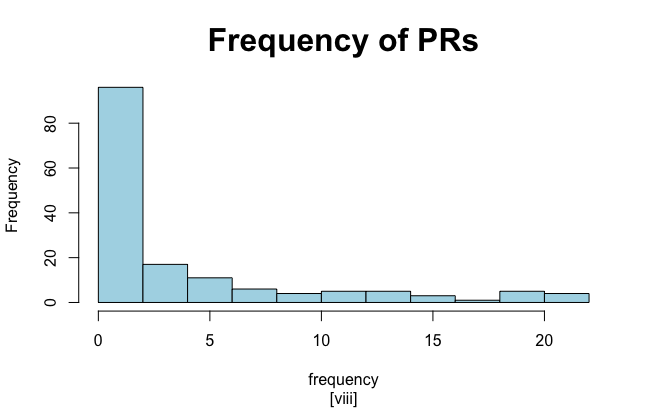}
    \small
    \label{fig:histogram8}
    
    \caption{\textbf{Histogram distribution of our metrics data. Note: \textit{presence of hard forks} has binary data. The red arrows point towards the hypothesized higher efficiency for the metric. Note: \textit{Dimensionality of commits} and \textit{Frequency of PRs} has no arrows as we do not hypothesize any direction of efficiency but only a presence of correlation with the outcome of a project.}}
    \label{fig:histogram}
\end{figure}

After all, values are gathered and calculated, the metrics are cleaned and analyzed using R. Using box plots, as shown Fig.\ref{fig:boxplot}, we inspected the data variability and removed the outliers that are above or below the 1.5 interquartile range. In Fig.\ref{fig:histogram}, we show the data distributions of the collected metrics in our dataset after the cleaning process.

Next, we use logistic regression to assess the significance of each predictor since logistic regression is more suitable for a value with a binomial property like the success of a project. It will tell us whether the predictor is significant or not. If a predictor is significant we can also use the coefficient to assess the scale of the predictor whether it has a high influence or not. Then we use backward stepwise regression, using the stepAIC function in R, that seeks the model with the least Akaike Information Criterion(AIC) possible to determine the predictor(s) among all metrics to find the most relevant variable to the success of a project. Lastly, we crosscheck how well the model fits using McFadden’s pseudo-R$^{2}$  measure.

\section{Results}
In this section, we discuss the results from the quantitative data analysis. Using logistic regression and backward stepwise regression, we examined the outcomes corresponding to our research questions. 

\subsection{Statistical Analysis}
We used logistic regression to test each metric for its significance on the success or failure of a project based on the graduated and retired status in ASFI dataset \cite{dataset}. Table [\ref{table:logistic_regression}] shows the summary of the logistic regression model, where each metric was separately considered for its significance. And, table [\ref{table:r2}] shows the summary of McFadden’s pseudo-R$^{2}$ measure for all of the metrics. \\

\begin{centering}
\begin{table}
\small
\caption{\textbf{Logistic Regression Summary}}
\setlength\tabcolsep{2pt}
\begin{tabular}{l*{2}{c}r}
\textbf{Coefficients} & \textbf{Estimate Std.} & \textbf{p-value} \\
\hline
(Intercept) & 0.88917 & 0.000258***   \\
Frequency Index   & 0.03136 & 0.554512   \\
(Intercept) & 0.41228 & 0.495   \\
Dimensionality Index   & 0.01685 & 0.283   \\
(Intercept) & 2.6377 & 2.2e-05***   \\
Additive Contribution Index   & -17.3226 & 0.00768**    \\
(Intercept) & 1.7177 & 1.26e-06***   \\
Logic Coupling Index   & -151.6851 & 0.0434*    \\
(Intercept) & 0.5952 & 0.0157*   \\
Central Management Index   & 13.1342 & 0.0209*    \\
(Intercept) & 1.5773 & 3.93e-16***   \\
Pre-communication Index   & 4.0689 & 0.227   \\
(Intercept) & 1.6081 & 2e-16***  \\
Ratio of Duplicate PRs    & 324.1628 & 0.25   \\
(Intercept) & 0.4447 & 0.08262   \\
Hard forks   & 1.3318 & 0.00132   \\
\hline
Signif. codes:  0 ‘***’ 0.001 & ‘**’ 0.01 ‘*’ & 0.05 ‘.’ 0.1 ‘ ’ 1\\
\hline
\end{tabular}\\
\label{table:logistic_regression}
\end{table}
\end{centering}
\vspace*{-0.5em}

\textbf{RQ\ref{rq:first}: Is there a difference in the number of duplicate PRs and the existence of hard forks between successful and unsuccessful projects? [H\ref{hypo:first}] [H\ref{hypo:second}]}\\
The ratio of PRs raised having redundant development is not a significant variable with a p-value of 0.244, as shown in the table [\ref{table:logistic_regression}] above. But the presence of hard forks [H\ref{hypo:second}] is indeed significant having a p-value of 0.00132. We, therefore, conclude that the number of duplicate PRs having redundant development does not correlate to the success of a project but the presence of hard forks does. We did not find enough evidence to accept [H\ref{hypo:first}]. For the [H\ref{hypo:second}], we found an interesting observation. The number of hard forks has a positive coefficient in our regression model, indicating that the number of hard forks in a project increases the probability of success of a project. But, according to our hypotheses [H\ref{hypo:second}], there should be relatively fewer hard forks for a successful project.\\

\textbf{RQ\ref{rq:second}: Does the frequency of pull requests (PRs) raised to the main project have anything to do with the success of a project? [H\ref{hypo:eigth}]}\\ 
To test our hypothesis [H\ref{hypo:eigth}], we built a logistic regression model keeping frequency index as an independent variable. From table [\ref{table:logistic_regression}] we found that the numbers of PR's raised within a time frame have a p-value of 0.492. This is greater than 0.05, which implies that it is not a significant indicator in determining the success of a project. Thus, we reject our hypothesis [H\ref{hypo:eigth}] since we didn't find enough evidence to accept it. In short, we found that the frequency of pull requests does not correlate with the success of a project.\\

\textbf{RQ\ref{rq:third}:  Is the size of the commits pushed to the main branch relevant to the success of a project? [H\ref{hypo:seventh}]}\\
The size of the commits is another one of the additional metrics that we believed to have no relationship with the project success. The dimensionality index [H\ref{hypo:seventh}] in our regression model resulted in a p-value of 0.145 which is greater than the expected value of 0.05. This indicates that it is not a significant variable relevant to the success of a project. Therefore, we accept the null hypothesis [H\ref{hypo:seventh}] to answer our third research question, i.e. the size of commits does not have any correlation with the success of a project.\\

\begin{centering}
\begin{table}
\caption{\textbf{R$^{2}$ value}}
\small
\setlength\tabcolsep{3pt}
\begin{tabular}{l*{2}{c}r}
\textbf{Metrics} & \textbf{McFadden’s pseudo-R$^{2}$} \\
\hline
Frequency Index   & 0.00243885   \\
Dimensionality Index   & 0.01388824  \\
Additive Contribution Index   & 0.04913989   \\
Logic Coupling Index   & 0.02475378  \\
Central Management Index   & 0.06018577  \\
Pre-communication Index   & 0.1994757  \\
Ratio of Duplicate PRs    & 0.01255469  \\
Hard forks   & 0.07834034 \\
\hline
\end{tabular}\\
\label{table:r2}
\end{table}
\end{centering}

\textbf{RQ\ref{rq:fourth}:  Is modularity of a project characteristic of it's success? [H\ref{hypo:third}, H\ref{hypo:fourth}]} \\
We were looking for the correlation of modular architecture of a project with its success. In our methodology, the modularity of a project is quantified using two metrics, logic-coupling index, and additive contribution index. From table [\ref{table:logistic_regression}], both additive contribution index and logic coupling index has a p-value lower than 0.05, which implied that they are significant metrics to determine the success of a project. The coefficient of the logic coupling index is negative, conforming to our hypothesis [H\ref{hypo:third}] that a successful project should not have logic coupling in multiple files across the project. However, in contrast to our hypothesis [H\ref{hypo:fourth}], the coefficient for the additive contribution index was negative. It implied that more files added while developing software may increase the possibility of failure to the project. McFadden’s pseudo-R$^{2}$ for additive contribution index and logic coupling index are 0.05 and 0.03 consecutively which indicated that the additive contribution index is a better predictor compared to Logic coupling in terms of determining the success of a project. Therefore, we can answer our research question that modularity is significant to the success of a project. The less logic-coupling and fewer added files in the project signify a better chance that the project will be a success.\\

\textbf{{RQ\ref{rq:fifth}: Is coordination between the developers and the core team of a project for working on an issue or new feature relevant to its success? [H\ref{hypo:fifth}, H\ref{hypo:sixth}]}} \\
The coordination between developers and the core team was measured using two metrics, the pre-communication index, and the central management index. As shown in the table [\ref{table:logistic_regression}], the p-value for the central management index is 0.0209, signifying a correlation to the success of a project. The positive coefficient for this metric confirmed our hypothesis [H\ref{hypo:fifth}] and provided us with enough evidence to accept it. For the pre-communication index, the p-value suggested that it is not a significant predictor. Hence, we reject the hypothesis [H\ref{hypo:sixth}]. Furthermore, McFadden’s pseudo-R$^{2}$ for central management index of 0.05 implied that central management is a better model to predict the success of a project. To summarize, we infer from our results that the interaction between the contributors of a project before a PR is important for its success. Whereas, declaration of commitment to an issue is not necessarily important for its success. 


\begin{centering}
\begin{table}
\caption{\textbf{stepAIC Summary}}
\small
\setlength\tabcolsep{3pt}
\begin{tabular}{l*{2}{c}r}
\textbf{Coefficients} & \textbf{Estimate Std.}  & \textbf{p-value} \\
\hline
(Intercept) & 1.009589 & 0.1071\\
Dimensionality   & 0.007572  & 0.3596 \\
Additive Index   & -8.568041  & 0.0569 \\
Hard forks   & 1.083335  & 0.0144*  \\
\hline
Signif. codes:   0 ‘***’ & 0.001  ‘**’ 0.01 ‘*’ & 0.05 ‘.’ 0.1 ‘ ’ 1\\
\hline
\end{tabular}\\
\label{table:stepAIC}
\end{table}
\end{centering}

\subsection{Backward  Stepwise  Regression}
After we investigated the significance of the metrics individually to a project's success using a logistic regression model, we performed a backward stepwise selection analysis, using stepAIC function in R, on all of our metrics combined. The results, as shown in the table [\ref{table:stepAIC}], confirms that two of the metrics which are additive contribution index and presence of hard forks are significant predictors for a project's success. The central management index and logic coupling index, even though correlated individually, are not considered as significant predictors in the stepAIC analysis.

\subsection{Conclusion} 
From our findings, we answer our main research question, that \textbf{not all forking practices we examined to lead to a project's success.} Every measure of modularity (logic coupling or additive contribution), which is ardently supported by the software community, does not increase the odds of the success of a project. The same findings were observed in the case of the management of a project. Understanding the trajectory of the project's future goals by contributors and maintainers (central management) is more crucial than communication about who will be working on a particular task (pre-communication). Furthermore, the presence of hard forks in a project, as well as the rate at which contributors attempt to incorporate modifications to a project, have no negative impact on its success. The size of the commits (dimensionality) is not a determining factor as it does not vary significantly between successful and retired projects. Since multiple duplicate pull requests (PRs) are rare in projects in the ASFI dataset, their presence isn't a significant factor that is correlated to a project's failure.
\section{Discussion}

In this section, we present the main findings derived from the results of our research. We also mention the strengths and limitations of our study, including threats to validity. Finally, we discuss future work that can be done based on our findings.

\subsection{Main Findings}

Based on the results, we present the following findings based on our metrics data.

\textbf{Redundant development:} In software engineering, redundant work is seen as a negative attribute in project development. We hypothesized that the ratio of duplicate PRs should be low [H\ref{hypo:first}] for a successful project as it indicates that people are doing a similar task, which wastes development time and resources \cite{produce_oss, what_the_fork}. Our model doesn’t suggest that there’s adequate evidence that the ratio of duplicate PRs has a significant impact on project success. We suspect this happens because duplicates are rare in most projects in ASFI \cite{what_the_fork}. This idea is further supported by the fact that we get sparse data for this metric. The number of projects with duplicate PRs was extremely low, and the quantity of duplicate PRs was minimal when compared to the overall number of PRs closed.

\textbf{Hard Forks:} We also hypothesized that the presence of hard forks is not a desired quality for a successful project [H\ref{hypo:second}] as it leads to fragmented communities \cite{social_forking, what_the_fork}. However, we discovered that hard forks are not a "negative" attribute that will cause a project to fail based on the results of our statistical analysis. We infer this because the coefficient for this metric was positive implying that the presence of hard forks increases the likelihood of a project becoming successful. We further believe that since the project has been successful,  many developers wish to use it as their base and contribute to it, or change it to fit their needs. In addition, we find that regardless of the presence of hard forks, successful projects with solid central management and a clear vision of the project's trajectory draw contributors to the main project and keep it alive.

\textbf{Modularity:} The importance of modularity in software engineering projects is well understood as it facilitates software evolution and provides collaboration \cite{how_to_invent, produce_oss, on_the_criteria}. Both of the modularity measures, i.e. logic coupling index, and additive contribution index, have been found as significant factors in determining success. The coefficient of additive contribution index, however, in the logistic regression table  [\ref{table:logistic_regression}] suggests otherwise [H\ref{hypo:fourth}]. From our interpretation, both measures of modularity do not signify the direction of a project, i.e. whether it is going to be successful or not. To be exact, more files added than being modified during the development does not mean the project is going to be successful.

\textbf{Coordination:} Even though forking has given the freedom to the developers to create a local copy without any intervention from the maintainers and contribute back when the development is done \cite{trans_and_colla}, coordination is associated with significant improvements to the efficiency of an open-source community. In our research, the result supports the hypothesis on the importance of central coordination. We have the evidence to accept [H\ref{hypo:fifth}] which indicates that the central management index is highly influential to the success of a project. On the contrary, we do not have enough evidence to confirm the significance of the pre-communication index. A possible explanation to why we do not have enough evidence is that the contributors could have used a communication channel other than GitHub Issue and PR section. Furthermore, we only examined the recent 150 PRs and Issues discussions associated with the PRs to reduce the exorbitant time of fetching the entire data over the GitHub API. Nevertheless, from the open-source communication over GitHub, it is implied from our analysis that pre-communication was not significant for the success of the ASFI projects. From the result of our research, we can conclude that it is more important to have central communication on "what" to do with the project rather than "who" would do the part.

\textbf{Dimensionality and frequency of contributions:} We also formulated two of the metrics which measure the frequency of the PRs raised and the size of the commits pushed by the developers. Our model suggests that the size of the commits does not correlate with the success of a project which aligns with our [H\ref{hypo:seventh}]. Kolassa et al. in their exploration \cite{commit_size} also mentions that indeed the scale of a project or the number of developers making contributions have no significant influence on the commit size, i.e. it more or less remains the same. However, our AIC model results, as seen in table [\ref{table:stepAIC}] indicates the dimensionality metric as one of the significant predictors, which is surprising as we didn't find enough evidence from our model to reject the [H\ref{hypo:seventh}]. We need further research and analysis to check why the metric came out significant in AIC. Our model also did not give us any significant evidence to accept the [H\ref{hypo:eigth}]. We assume this happens as the core team isn't concerned about the quantity of PRs but about the ones targeting a specific issue, tracked by the issue trackers in the version control system. So even though the rate of contribution across the project is high, good central coordination keeps the focus of the maintainers on the PRs achieving the goals as envisioned by the maintainers. 

\subsection{Strengths and Limitations}

Prior research on the sustainability of projects did not have any objective way to declare a project as a success or failure. In our study, we examined the ASFI dataset, which includes specific labels for the status of each project. As a result, we were able to link the project's qualities to objectively specified labels. Previous works \cite{what_the_fork, perspective_on_forking, social_coding_forking, identifying_redundancies} have looked into efficient forking practices to foster contributions, but they didn't study whether self-sustaining projects use those efficient forking practices. We investigated this through our study, where we test our hypotheses and formulate metrics for assessing the correlation between forking practices and the sustainability of a project. Additionally, our project features robust python scripts for mining GitHub data, which can be used to enhance the project list.

\textbf{Threats to validity:}
One of the main drawbacks of using the ASFI dataset is generalizing the research implications beyond the ASF community. The list of projects is of a relatively small scale, with hundreds of projects only. There is also limited diversity as we see an inherent imbalance between graduated vs. retired projects. There may also be a presence of systematic bias in our data, meaning what we measure is not the likelihood of getting that metric data. With automatic detection, false-positive errors are inevitable, which means that some pull requests (PRs) reported as duplicate pull requests (PRs) aren't duplicates when their context is studied thoroughly. Due to limited access issues in the GitHub API, we did not identify external contributors. As a result, some of the forks may be incorrectly categorized as hard forks only based on the number of PRs in that fork.

Furthermore, developers may have opted to communicate outside of GitHub Pull Request and Issue sections. We might have introduced some threat to internal validity as we only calculate our metrics based on the information available on the project's GitHub page and not external chats or emails. The pre-communication index required more than 100,000 requests over the GitHub API due to the hierarchical nature of the metric. To reduce the calculation time of this metric, we only consider the first 5 pages or 150 recent Pull Requests per project.

The assumptions mentioned for redundant development are somewhat contrary to Apache Software community guidelines as communication is of utmost importance in ASF projects to cause redundancy. However, it is not hard to imagine these scenarios where that data may not reflect as we think it is.

\subsection{Future Work}

We performed backward stepwise regression using all of our metrics as a starter model. As shown in the table [\ref{table:stepAIC}], the metrics that are left are dimensionality index, additive contribution index, and presence of hard forks. The dimensionality index, however, was not a significant predictor when considered individually and hence needs further research. A machine learning model can also be created to forecast the probability of success of a project. For some of our metrics, the quality of data collected can be improved by incorporating data about external contributors and external communication channels. Also, for metrics relying heavily on GitHub APIs, we capped the number of pages fetched considering the rate limit policy of GitHub and the time constraints. We can refine them by using alternate GitHub mining sources such as GHTorrent. Finally, we can extend our research beyond the ASFI dataset to a larger dataset with a variety of projects to generalize the findings of our research to the broader OSS community.

\section{Team member and Attestation}
Anurag Dhasmana, Arindaam Roy, Divjeet Singh Jas, Kiranpreet Kaur, Pinn Prugsanapan \\
\emph{*All team members have read and contributed equally to this research.}

\small{
\bibliographystyle{unsrt}
\bibliography{egbib}
}

\end{document}